\RequirePackage{silence}
\PassOptionsToPackage{unicode, pdfprintscaling=None, breaklinks, colorlinks, allcolors=blue!50!black}{hyperref}
\PassOptionsToPackage{usenames}{xcolor}

\documentclass[aps, 11pt, prd,  preprintnumbers, floatfix, unsortedaddress, superscriptaddress, longbibliography, nofootinbib,twocolumn,tightenlines]{revtex4-2}
\ifx\pdfsuppresswarningpagegroup\undefined\else\pdfsuppresswarningpagegroup=1\fi\relax
\newif\ifdomakeglossary

\usepackage{orcidlink,multirow,siunitx,tikz,graphics,graphicx,times,latexsym,mathtools,%
            amsmath,amssymb,amsbsy,amsfonts,array,bm,mathrsfs,inconsolata,cancel,%
            makecell,microtype,enumitem,FiraSans,FiraMono,silence,makeidx,enumerate,%
            enumitem,tabularx,color,pslatex,fancyhdr,wrapfig,float,mdwlist,pgfgantt,%
            multirow,slashed,ucs,soul,subfigure,caption,braket}
\usepackage[T1]{fontenc}
\usepackage{parskip}
\usepackage[usenames]{xcolor}
\usepackage[compact]{titlesec}
\usepackage{mdwlist}
\usepackage[margin=1in]{geometry}
\usepackage{breakurl}
\usepackage{ctable}
\ifdomakeglossary
 \usepackage[acronyms, nopostdot, style=super, nonumberlist, toc]{glossaries}
\else
 \usepackage[acronyms, nohypertypes={acronym}, nopostdot, style=super, nonumberlist, toc]{glossaries}
\fi
\usepackage[unicode, pdfprintscaling=None, breaklinks, colorlinks, allcolors=blue!50!black]{hyperref}
\usepackage{pdfcomment,hypernat}
\usepackage[normalem]{ulem}
\usepackage[capitalise, nameinlink]{cleveref}

\setlist[itemize]{leftmargin=4.0mm}
\usetikzlibrary{decorations.markings, arrows, positioning, calc}
\graphicspath{{./fig/}}
\glsenableentrycount
\ifdomakeglossary\makenoidxglossaries\fi

\DeclareMathOperator\Tr{Tr}

\newcommand\footnotepunct[2]{{\edef\tmp{\spacefactor=\the\spacefactor\relax#2}%
    \toks0={#1}%
    \hbox to 0pt{\tmp
      \edef\tmp{\noexpand\footnote{\the\toks0}%
        \spacefactor=\the\spacefactor\relax}%
      \hss\expandafter}%
    \tmp}}

\makeatletter
\newcommand\showtitleinbib{{\escapechar=`\\ \immediate\write\@auxout{%
\csname citation{REVTEX42Control}\endcsname^^J%
\csname citation{apsrev42Control}\endcsname
}}}
\makeatother

\newcommand\breakableslash{/\hspace{0pt}}
\newcommand\breakablestop{.\hspace{0pt}}
{\catcode`\/=\active \catcode`\.=\active
 \toks0={\let/\breakableslash\let.\breakablestop}
 \expandafter}\the\toks0
\let\orighref\href
\newcommand\passtohref[2]{\orighref{#1}{#2}}

\newif\ifshowchanges\showchangestrue
\ifshowchanges
\usepackage{soul}\setul{1ex}{0.25ex}
\fi
\newcommand\defeditcmd[2]{%
   \ifshowchanges
      \newcommand#1[2][]{%
        \ifx\undefined##1\undefined
        \else\setstcolor{#2}\st{##1}\fi
        {\color{#2}##2}}%
  \else\newcommand#1[2][]{{##2}}\fi}

\defeditcmd\TB{green!60!black}

\makeatletter
\@ifundefined{texorpdfstring}
             {\newcommand\texorpdfstring[2]{#1}}
             {}
\@ifundefined{unichar}
             {\newcommand\myunichardef[3]{\expandafter\providecommand\csname text#1\endcsname
                 {#2}}}
             {\newcommand\myunichardef[3]{\expandafter\providecommand\csname text#1\endcsname
                 {\unichar{"#3}}}}

\makeatother
\myunichardef{composetilde}{\string~}{0303}
\myunichardef{composemacron}{bar}{0305}
\myunichardef{subscriptthree}{\string_3}{2083}
\myunichardef{superscripttwo}{\string^2}{00B2}

\newcommand\ac[1]{\gls{#1}}

\makeatletter
\ifx\switch@longtable\undefined\else{\catcode`\*=11
\def\tmp{\expandafter\def\expandafter\switch@longtable\expandafter{\switch@longtable
\let\longtable*\@undefined\let\endlongtable*\undefined}}\expandafter}\tmp\fi
\makeatother

\newacronym{qcd}{QCD}{quantum chromodynamics}
\newacronym{cp}{CP}{charge-conjugation--parity}
\newacronym{cpv}{CPV}{{\ac{cp}} violation}
\newacronym{sm}{SM}{standard model of paricle physics}
\newacronym{bsm}{BSM}{beyond the {\ac{sm}}}

\begin{document}

\title{Comment on the claim of physical irrelevance of the topological term}
\author{Tanmoy Bhattacharya\,\orcidlink{0000-0002-1060-652X}}
\email{tanmoy@lanl.gov}
\affiliation{Los Alamos National Laboratory, Los Alamos, New Mexico 87545, USA}
\preprint{LA-UR-22-29996}

\begin{abstract}
We argue that the claim~\cite{Ai:2020ptm} of the absence of \ac{cp} violating effects due to a topological term in \ac{qcd} is based upon a misunderstanding, and the standard results in the field are correct.
\end{abstract}

\maketitle

\section{Introduction}
Our matter dominated universe is a profound mystery for the \ac{sm}---there is no known mechanism that can segregate baryons and antibaryons to prevent them from almost compete annihilation in a symmetric universe, a large enough initial baryon asymmetry to survive the inflationary epoch would overclose the universe, and dynamical creation of the asymmetry needs violation of the approximate \ac{cp} symmetry of the standard model. As a result, it is there is a large experimental, and concomitant theoretical, effort looking from \ac{cpv} arising from theories \ac{bsm}.

One such source of \ac{cpv} is due to a dimension-4 operator called \ac{qcd} topological \(\Theta\)-term, \(\Theta Q \equiv \frac\Theta{16\pi^2} \int \Tr G \wedge G\). The corresponding topological charge density is \(q \equiv \Tr G \wedge G\), or \(\Tr G \cdot \tilde G \) in components, where the trace is over the color indices, \(G \equiv d A + A \wedge A\), or \(G_{\mu\nu} = \partial_\mu A_\nu - \partial_\nu A_\mu + [A_\mu, A_\nu]\) in components, is the gauge-field adjoint two-form which is the `curvature' of the principal connection one-form, \(A\), giving the gauge potential, and \(\tilde G \equiv *G\), or, in components, \(\tilde G^{\mu\nu} \equiv \frac12\epsilon^{\kappa\lambda\mu\nu}G_{\kappa\lambda}\), is the dual gauge field.  This density is topological since locally, it is an exact form, \(q = d K\), {\itshape i.e.,} the exterior derivative of a 3-form \(K= dA \wedge A + \frac23 A\wedge A\wedge A\).  As a result, if the gauge potential \(A\) is a trivializable, i.e., if it can be represented as a smooth algebra valued one-form, then \(Q = \oint K\) is  surface integral or zero if there are no boundaries.  The topological nature is manifested in the fact that in spaces with no boundaries, there may be gauge field configurations that do not arise from trivializable connections, in other words, they can be written as smooth one-forms only on open subsets and on the overlap of these subsets, they are related by a gauge transformation, so that the field strength is still unique in this overlap.

\section{Specific Rebuttal}
In a recent publication, \citet{Ai:2020ptm} claim that if the calculations are carried out properly, a \ac{cpv} cannot be introduced by a topological term in the action of \ac{qcd}. This, of course, goes against standard understanding of topological effects in non-Abelian theories~\cite{Callan:1976je}, and, if true, would lead to an automatic solution to the so-called strong \ac{cp} problem and possibly remove much of the interest in axion physics~\cite{Peccei:1977hh,Peccei:1977ur}.  It would also doom the efforts of putting limits on the so-called \ac{qcd} \(\theta\)-term\footnote{There is an independent claim~\cite{Nakamura:2021meh} of numerical evidence that such a term in the action kills confinement at distances greater than \(1/O(\theta^2\Lambda_{\rm QCD})\sim O({\rm km})\) for \(\theta\sim 10^{-9}\), and so for confinement to be maintained, \(\theta\) must dynamically flow to 0 in the deep infrared; even though no evidence for such a flow was demonstrated unless confinement is \emph{assumed.} Nothing we say here relates to that claim except to note that such arguments are independent of the discussions here, and even if it could be convincingly demonstrated, the phenomenological consequences of any such effect is probably negligible for strong interaction physics at distances of a few femtometers.} by putting limits on the neutron electric dipole moment~\cite{Baker:2006ts}.  Here we show this is not the case---the so-called \ac{qcd} \(\theta\)-term does violate \ac{cp} in physical processes exactly as understood currently.

Let us recap the standard understanding in the path integral formulation.  If we add a term  \(\frac\theta{16\pi^2} \int \Tr G \wedge G\) to the action, where \(G \equiv dA + A\wedge A\) is the algebra-valued gauge-field two-form with \(A\) the algebra-valued gauge connection one-form, we expect to break the \ac{cp}-symmetry of the action.  This term, however, is a total derivative and, if \(A\) is single-valued, can be written as a surface integral, \(Q\equiv\oint (K \equiv dA \wedge A + \frac23 A \wedge A \wedge A)\) over the space-time boundary of the physical region if the region has a boundary\footnote{To avoid confusion, we emphasize that we have not assumed infinite volumes or any specific boundary condition in making this statement.} and zero otherwise. One usually studies the theory while demanding that the gauge fields \(G\) fall off as one moves outside a finite space-time region.\footnote{One usually claims this as a corollary of the configuration having finite action. In fact, however, one always has finite action configurations where the gauge fields are nonzero in a region of finite extent near the boundary. We can choose not to consider them because of cluster decomposition: as the boundary moves infinitely far away, the value of field variables at infinity should have no impact on local physics, so the conventional choice of setting \(G\to0\) at the boundary is just a convenient choice.  Note that our discussion has nothing to do with arguments about the validity of saddle point approximation brought up in Supplemental section S3.2 of \citet{Ai:2020ptm}, since we are not interested in such approximations, only in matters of principles raised there.}  In Abelian gauge theories without magnetic monopoles, \(A\) can indeed be chosen to be single valued and all such finite action configurations  satisfy \(K\to0\) sufficiently fast at large distances, so that the effect of the topological term vanishes if we consider either infinite volumes or space-times without a boundary.  As a result, physics is independent of \(\theta\), and all physical effects, including \ac{cpv}, arising from it vanish.  In non-Abelian gauge theories, however, this result breaks down: even classically, there are finite action configurations called instantons, such that they have localized gauge fields but have a nonzero integral value of \(Q\), for every integer, at arbitrarily large volumes. This arises since the gauge connection \(A\), by its nature, is not necessarily single-valued as a function: it can be defined as a function in open space-time regions, with gauge transformations relating its values on the overlap of these regions.\footnote{In passing, we note that these gauge configurations, though intrinsically nonperturbative, make a nonzero contribution to physical matrix elements, as can be checked explicitly with a gauge invariant regularization like the lattice---an issue that makes the arguments following Eq.~(5) of \citet{Yamanaka:2022vdt} invalid.} The topology of these overlaps allows \(Q\) to take only integral values when there is no boundary contribution.\footnote{As noted in section 2 of \citet{Ai:2020ptm}, we can shrink the regions to neighborhood of points and study what are called `singular gauges'.  The issue is not the language we speak: rather it is that the topological charge gets an integer bulk contribution in addition to any surface contribution.} As a result, the addition of this term to the action has non-trivial effect on the physics---one notes, however, that the physics derived from the theories with \(\theta\) and \(\theta+2n\pi\) for any integer \(n\) become the same as the volume increases to infinity, but the results do depend on \(\theta\in[0,2\pi\mathclose[\).  Nothing that is discussed in \citet{Ai:2020ptm} addresses this straightforward calculation that shows that we can, indeed, study the effect of the \(\theta\)-term on physics in a finite volume.\looseness-1

In their Supplemental section S3.2, they do discuss doing the calculation in finite boxes.  They admit there that if one did this calculation with a boundary condition specifying zero gauge field, i.e., pure gauge potential, they would get the standard result.  They object, however, there is no reason to assume a zero gauge field boundary condition in finite volume, and we agree; it is more natural to use open boundary condition when studying a finite physical region of space-time as they advocate.  They, however, do not carry out this calculation, simply claiming that such boundary conditions do not lead to quantization of the topological charge. Our point in this rebuttal is that it does not matter for physical quantities calculated far away from the boundary. In theories with a mass gap, the boundary conditions only give corrections exponential in the distance to the boundary when the entire system is scaled up in size: so one can indeed replace the open boundary condition with the standard zero-field boundary condition if one is interested in the large volume limit of locally-defined quantities.\footnote{We ignore complications if multiparticle states are being studied since they do have a polynomial volume dependence in their phase space.} Since in the latter case the charge is indeed quantized, all physical effects of charge quantization, such as \(2\pi\) periodicity in the Lagrangian parameter \(\theta\), must appear for all physical quantities calculated far away from the boundary, even with open boundary conditions, with exponentially suppressed corrections.  Nothing in the discussion provided by \citet{Ai:2020ptm} refutes this, and it is to be noticed in this respect that lattice calculations are routinely done with very different boundary conditions (e.g., the usual periodic, C-periodic~\cite{Kronfeld:1990qu}, and open~\cite{Luscher:2011kk,Gattringer:2021xrb} on the hypertoroidal lattice) and no boundary condition dependence has ever been observed in the large volume limit.

Let us now discuss the comments about the nature of the topological sectors being made in \citet{Ai:2020ptm}.  One can indeed start with the Hilbert space of a non-Abelian gauge theory written in the gauge field-basis, i.e., we can define our basis states to be such that \(\hat A^a_\mu(x) \ket{A} = A^a_\mu(x) \ket{A}\), where \(A^a_\mu\) are functions\footnote{Since all SU(3) principal bundles are trivializable in this case, i.e., there is no nontrivial topology to worry about when considering gauge connections in 3-space, we can just consider functions instead of having to deal with sections of bundles.  Note, however, that the space of gauge transformations, or equivalently the space of gauge connections equivalent to the trivial \(A_\mu=0\), still has a non-trivial topology, and hence, there are large gauge transformations, that cannot be continuously deformed to the identity.} of the space-coordinate \(x\).  One knows that this Hilbert space is too large: most of these states are not physical, and one needs to consider the subspace of `gauge-invariant' states. Since the states are invariant under gauge transformations only up to a phase, each of them forms a one-dimensional representation of the gauge group, and it is these representations that are parameterized by a real number \(\theta\in[0,2\pi)\).  In particular, we can generate the states in these sectors as \(\ket{A}^\theta_{\rm ph} = \sum_n e^{in\theta} {\cal G}^n \ket{A}_{\rm ph}\), where \({\cal G}\) is a generator of the group \(\mathbb Z\) of large gauge transformations, and the physical states are gauge invariant under gauge transformations connected to the identity, i.e., \(\ket{A}^g_{\rm ph}=\ket{A}_{\rm ph}\) for all small gauge transformations \(g\). This is the generalization of Supplementary Eq.~S66 of \citet{Ai:2020ptm}, except pointing out that these sectors exist not only for the configurations that are pure gauge, rather they exist for any physical state---the important thing is only the group structure \(\mathbb Z\) of the large gauge transformations, or, strictly speaking, the homotopy group of gauge transformations.\footnote{As a technical point, one could ask how this plays out if we discretize space-time on a lattice: after all, there is no topology on such a space. This is clear by noticing that any `instanton' can be smoothly shrunk to lie in an arbitrarily small region in the void between the lattice sites, and thus disappear in the limit. The solution turns out to be that if the action density is small everywhere, topological sectors appear on the lattice in exactly the same way as in the continuum~\cite{Luscher:1981zq,Phillips:1986qd}, and near the continuum limit defined by the critical point at zero gauge coupling, the configurations which have higher action densities than this cutoff anywhere on the lattice make no contribution. as described in Section 4 of \citet{Luscher:2010we}, and the irrelevance of the exact definition of the topological charge for physical quantities calculated on the lattice is demonstrated in \citet{Alexandrou:2017hqw}.} In fact, actual lattice calculations on finite volumes show that the topological charge becomes close to an integer if defined in terms of the Wilson-flowed gauge fields for importance-sampled configurations contributing to the path integral, see, for example, Figure 3 of \citet{Bhattacharya:2021lol}.\footnote{These calculations were done on a hypertoroidal boundary conditions.  The integerness of the topological charge under this boundary condition is discussed in \citet{vanBaal:1982ag}.}

It is easy to see that gauge invariance implies that the different values of \(\theta\) give superselection sectors: the  matrix element between any state belonging to one sector and any state in a different sector is identically zero, i.e., \({}^{\ \theta'}_{\rm ph}\!\bra{A'} O \ket{A}^\theta_{\rm ph} = \delta(\theta-\theta') \sum_{\Delta n} e^{i\theta\Delta n} \;{}_{\rm ph}\!\bra{A'} O {\cal G}^{\Delta n}\ket{A}_{\rm ph} \) for all gauge invariant operators \(O\).  As a result, we can choose to consider the physics in only one such sector,\footnote{And, by the requirement of the cyclicity of the vacuum, we must do so.} in which case it is convenient to shift the physics of the \(\theta\) into the action.

Given this discussion, it is clear that there is, in principle, no barrier to calculating physical correlation functions at a finite value of \(\theta\) even in finite volumes, even without imposing a Dirichlet boundary condition on the gauge fields on the lattice.

With this, one can go back and understand the infinite volume limit.  Like any extensive quantity, the instanton density grows with the volume.  In fact, the susceptibility \(\chi_q\equiv\langle Q^2\rangle/V\) reaches a finite value at large volumes even if we do not choose a definition of \(Q\) that is an integer on the lattice~\cite{Bhattacharya:2021lol}.  In fact, the sum over all topological sectors that give rise to the physical effects of \(\theta\) are dominated by configurations that have ever larger topological charges as the volume is increased; any simulation at finite topological charge fails to see this.

In fact, one can ask what is the quantity that one obtains if the simulation is done at finite topological charge.  Since large gauge transformations change the topological charge, correlation functions at fixed topological charge provide access to the Fourier transform of the result as a function of \(\theta\). Furthermore, an explicit calculation shows that a simulation with fixed topological charge \(Q\) at large volumes recover the physics of the \(\theta=0\) theory with finite volume corrections proportional to \(Q^2/V\)~\cite{Brower:2003yx}.\footnote{This reference provides explicit evidence that the fixed \(Q\) theory, because of the global constraint, is not even described by a local Hamiltonian at finite volumes. In particular, the two-point correlation functions in this theory do not have the usual spectral representation.} Since this result is independent of the \(\theta\) in the action---which at fixed \(Q\) merely multiplies the partition function by an irrelevant phase---incorrectly taking the infinite volume limit without allowing increasingly large topological charges only gives the physics of the \(\theta=0\) theory as observed in \citet{Ai:2020ptm}. If instead, we use a fixed, but volume-dependent, \(Q(V)\), increasing with volume, and chosen to be equal to \(\langle Q\rangle_\theta(V)\), the expectation value of the topological charge at the same volume and at given value of theta,\footnote{Which, as we stress, can be, and has been, calculated: see, for example, \citet{Guo:2015tla}.} then the usual arguments for the equivalence of canonical and grand canonical ensemble go through, and all physical quantities calculated at this value of \(Q(V)\) agree with the calculations at the given \(\theta\) as \(V\to\infty\).

\bibliographystyle{apsrev4-2-errat} 
\showtitleinbib
\bibliography{main}

\begin{thebibliography}{18}%
\makeatletter
\providecommand \@ifxundefined [1]{%
 \@ifx{#1\undefined}
}%
\providecommand \@ifnum [1]{%
 \ifnum #1\expandafter \@firstoftwo
 \else \expandafter \@secondoftwo
 \fi
}%
\providecommand \@ifx [1]{%
 \ifx #1\expandafter \@firstoftwo
 \else \expandafter \@secondoftwo
 \fi
}%
\providecommand \natexlab [1]{#1}%
\providecommand \enquote  [1]{``#1''}%
\providecommand \bibnamefont  [1]{#1}%
\providecommand \bibfnamefont [1]{#1}%
\providecommand \citenamefont [1]{#1}%
\providecommand \href@noop [0]{\@secondoftwo}%
\providecommand \href [0]{\begingroup \@sanitize@url \@href}%
\providecommand \@href[1]{\@@startlink{#1}\@@href}%
\providecommand \@@href[1]{\endgroup#1\@@endlink}%
\providecommand \@sanitize@url [0]{\catcode `\\12\catcode `\$12\catcode
  `\&12\catcode `\#12\catcode `\^12\catcode `\_12\catcode `\%12\relax}%
\providecommand \@@startlink[1]{}%
\providecommand \@@endlink[0]{}%
\providecommand \url  [0]{\begingroup\@sanitize@url \@url }%
\providecommand \@url [1]{\endgroup\@href {#1}{\urlprefix }}%
\providecommand \urlprefix  [0]{URL }%
\providecommand \Eprint [0]{\href }%
\providecommand \doibase [0]{https://doi.org/}%
\providecommand \selectlanguage [0]{\@gobble}%
\providecommand \bibinfo  [0]{\@secondoftwo}%
\providecommand \bibfield  [0]{\@secondoftwo}%
\providecommand \translation [1]{[#1]}%
\providecommand \BibitemOpen [0]{}%
\providecommand \bibitemStop [0]{}%
\providecommand \bibitemNoStop [0]{.\EOS\space}%
\providecommand \EOS [0]{\spacefactor3000\relax}%
\providecommand \BibitemShut  [1]{\csname bibitem#1\endcsname}%
\let\auto@bib@innerbib\@empty
\bibitem [{\citenamefont {Ai}\ \emph {et~al.}(2021)\citenamefont {Ai},
  \citenamefont {Cruz}, \citenamefont {Garbrecht},\ and\ \citenamefont
  {Tamarit}}]{Ai:2020ptm}%
  \BibitemOpen
  \bibfield  {author} {\bibinfo {author} {\bibfnamefont {W.-Y.}\ \bibnamefont
  {Ai}}, \bibinfo {author} {\bibfnamefont {J.~S.}\ \bibnamefont {Cruz}},
  \bibinfo {author} {\bibfnamefont {B.}~\bibnamefont {Garbrecht}},\ and\
  \bibinfo {author} {\bibfnamefont {C.}~\bibnamefont {Tamarit}},\ }\bibfield
  {title} {\bibinfo {title} {Consequences of the order of the limit of infinite
  spacetime volume and the sum over topological sectors for {CP} violation in
  the strong interactions},\ }\href
  {https://doi.org/10.1016/j.physletb.2021.136616} {\bibfield  {journal}
  {\bibinfo  {journal} {Phys. Lett. B}\ }\textbf {\bibinfo {volume} {822}},\
  \bibinfo {pages} {136616} (\bibinfo {year} {2021})},\ \Eprint
  {https://arxiv.org/abs/2001.07152} {arXiv:2001.07152 [hep-th]} \BibitemShut
  {NoStop}%
\bibitem [{\citenamefont {Callan}\ \emph {et~al.}(1976)\citenamefont {Callan},
  \citenamefont {Dashen},\ and\ \citenamefont {Gross}}]{Callan:1976je}%
  \BibitemOpen
  \bibfield  {author} {\bibinfo {author} {\bibfnamefont {C.~G.}\ \bibnamefont
  {Callan}, \bibfnamefont {Jr.}}, \bibinfo {author} {\bibfnamefont {R.~F.}\
  \bibnamefont {Dashen}},\ and\ \bibinfo {author} {\bibfnamefont {D.~J.}\
  \bibnamefont {Gross}},\ }\bibfield  {title} {\bibinfo {title} {The structure
  of the gauge theory vacuum},\ }\href
  {https://doi.org/10.1016/0370-2693(76)90277-X} {\bibfield  {journal}
  {\bibinfo  {journal} {Phys. Lett. B}\ }\textbf {\bibinfo {volume} {63}},\
  \bibinfo {pages} {334} (\bibinfo {year} {1976})}\BibitemShut {NoStop}%
\bibitem [{\citenamefont {Peccei}\ and\ \citenamefont
  {Quinn}(1977{\natexlab{a}})}]{Peccei:1977hh}%
  \BibitemOpen
  \bibfield  {author} {\bibinfo {author} {\bibfnamefont {R.~D.}\ \bibnamefont
  {Peccei}}\ and\ \bibinfo {author} {\bibfnamefont {H.~R.}\ \bibnamefont
  {Quinn}},\ }\bibfield  {title} {\bibinfo {title} {{CP} conservation in the
  presence of instantons},\ }\href
  {https://doi.org/10.1103/PhysRevLett.38.1440} {\bibfield  {journal} {\bibinfo
   {journal} {Phys. Rev. Lett.}\ }\textbf {\bibinfo {volume} {38}},\ \bibinfo
  {pages} {1440} (\bibinfo {year} {1977}{\natexlab{a}})}\BibitemShut {NoStop}%
\bibitem [{\citenamefont {Peccei}\ and\ \citenamefont
  {Quinn}(1977{\natexlab{b}})}]{Peccei:1977ur}%
  \BibitemOpen
  \bibfield  {author} {\bibinfo {author} {\bibfnamefont {R.~D.}\ \bibnamefont
  {Peccei}}\ and\ \bibinfo {author} {\bibfnamefont {H.~R.}\ \bibnamefont
  {Quinn}},\ }\bibfield  {title} {\bibinfo {title} {Constraints imposed by {CP}
  conservation in the presence of instantons},\ }\href
  {https://doi.org/10.1103/PhysRevD.16.1791} {\bibfield  {journal} {\bibinfo
  {journal} {Phys. Rev. D}\ }\textbf {\bibinfo {volume} {16}},\ \bibinfo
  {pages} {1791} (\bibinfo {year} {1977}{\natexlab{b}})}\BibitemShut {NoStop}%
\bibitem [{\citenamefont {Nakamura}\ and\ \citenamefont
  {Schierholz}(2021)}]{Nakamura:2021meh}%
  \BibitemOpen
  \bibfield  {author} {\bibinfo {author} {\bibfnamefont {Y.}~\bibnamefont
  {Nakamura}}\ and\ \bibinfo {author} {\bibfnamefont {G.}~\bibnamefont
  {Schierholz}},\ }\href@noop {} {\bibinfo {title} {The strong {CP} problem
  solved by itself due to long-distance vacuum effects}} (\bibinfo {year}
  {2021}),\ \Eprint {https://arxiv.org/abs/2106.11369} {arXiv:2106.11369
  [hep-ph]} \BibitemShut {NoStop}%
\bibitem [{\citenamefont {Baker}\ \emph {et~al.}(2006)\citenamefont {Baker}
  \emph {et~al.}}]{Baker:2006ts}%
  \BibitemOpen
  \bibfield  {author} {\bibinfo {author} {\bibfnamefont {C.~A.}\ \bibnamefont
  {Baker}} \emph {et~al.},\ }\bibfield  {title} {\bibinfo {title} {An improved
  experimental limit on the electric dipole moment of the neutron},\ }\href
  {https://doi.org/10.1103/PhysRevLett.97.131801} {\bibfield  {journal}
  {\bibinfo  {journal} {Phys. Rev. Lett.}\ }\textbf {\bibinfo {volume} {97}},\
  \bibinfo {pages} {131801} (\bibinfo {year} {2006})},\ \Eprint
  {https://arxiv.org/abs/hep-ex/0602020} {arXiv:hep-ex/0602020} \BibitemShut
  {NoStop}%
\bibitem [{\citenamefont {Yamanaka}(2022)}]{Yamanaka:2022vdt}%
  \BibitemOpen
  \bibfield  {author} {\bibinfo {author} {\bibfnamefont {N.}~\bibnamefont
  {Yamanaka}},\ }\href@noop {} {\bibinfo {title} {Unobservability of
  topological charge in nonabelian gauge theory}} (\bibinfo {year} {2022}),\
  \Eprint {https://arxiv.org/abs/2212.10994v1} {arXiv:2212.10994v1 [hep-th]}
  \BibitemShut {NoStop}%
\bibitem [{\citenamefont {Kronfeld}\ and\ \citenamefont
  {Wiese}(1991)}]{Kronfeld:1990qu}%
  \BibitemOpen
  \bibfield  {author} {\bibinfo {author} {\bibfnamefont {A.~S.}\ \bibnamefont
  {Kronfeld}}\ and\ \bibinfo {author} {\bibfnamefont {U.~J.}\ \bibnamefont
  {Wiese}},\ }\bibfield  {title} {\bibinfo {title} {{SU(N)} gauge theories with
  {C} periodic boundary conditions. 1. topological structure},\ }\href
  {https://doi.org/10.1016/0550-3213(91)90479-H} {\bibfield  {journal}
  {\bibinfo  {journal} {Nucl. Phys. B}\ }\textbf {\bibinfo {volume} {357}},\
  \bibinfo {pages} {521} (\bibinfo {year} {1991})}\BibitemShut {NoStop}%
\bibitem [{\citenamefont {Luscher}\ and\ \citenamefont
  {Schaefer}()}]{Luscher:2011kk}%
  \BibitemOpen
  \bibfield  {author} {\bibinfo {author} {\bibfnamefont {M.}~\bibnamefont
  {Luscher}}\ and\ \bibinfo {author} {\bibfnamefont {S.}~\bibnamefont
  {Schaefer}},\ }\bibfield  {title} {\bibinfo {title} {Lattice {QCD} without
  topology barriers},\ }\href {https://doi.org/10.1007/JHEP07(2011)036}
  {\bibfield  {journal} {\bibinfo  {journal} {JHEP}\ }\textbf {\bibinfo
  {volume} {2011}}\bibfield  {number} {\bibinfo  {number} { (07)},\ \bibinfo
  {pages} {036}},\ }\Eprint {https://arxiv.org/abs/1105.4749} {arXiv:1105.4749
  [hep-lat]} \BibitemShut {NoStop}%
\bibitem [{\citenamefont {Gattringer}\ and\ \citenamefont
  {Orasch}(2022)}]{Gattringer:2021xrb}%
  \BibitemOpen
  \bibfield  {author} {\bibinfo {author} {\bibfnamefont {C.}~\bibnamefont
  {Gattringer}}\ and\ \bibinfo {author} {\bibfnamefont {O.}~\bibnamefont
  {Orasch}},\ }\bibfield  {title} {\bibinfo {title} {Density of states approach
  for lattice field theory with topological terms},\ }\href
  {https://doi.org/10.22323/1.396.0158} {\bibfield  {journal} {\bibinfo
  {journal} {PoS}\ }\textbf {\bibinfo {volume} {LATTICE2021}},\ \bibinfo
  {pages} {158} (\bibinfo {year} {2022})},\ \Eprint
  {https://arxiv.org/abs/2111.09535} {arXiv:2111.09535 [hep-lat]} \BibitemShut
  {NoStop}%
\bibitem [{\citenamefont {Luscher}(1982)}]{Luscher:1981zq}%
  \BibitemOpen
  \bibfield  {author} {\bibinfo {author} {\bibfnamefont {M.}~\bibnamefont
  {Luscher}},\ }\bibfield  {title} {\bibinfo {title} {Topology of lattice gauge
  fields},\ }\href {https://doi.org/10.1007/BF02029132} {\bibfield  {journal}
  {\bibinfo  {journal} {Commun. Math. Phys.}\ }\textbf {\bibinfo {volume}
  {85}},\ \bibinfo {pages} {39} (\bibinfo {year} {1982})}\BibitemShut {NoStop}%
\bibitem [{\citenamefont {Phillips}\ and\ \citenamefont
  {Stone}(1986)}]{Phillips:1986qd}%
  \BibitemOpen
  \bibfield  {author} {\bibinfo {author} {\bibfnamefont {A.}~\bibnamefont
  {Phillips}}\ and\ \bibinfo {author} {\bibfnamefont {D.}~\bibnamefont
  {Stone}},\ }\bibfield  {title} {\bibinfo {title} {Lattice gauge fields,
  principal bundles and the calculation of topological charge},\ }\href
  {https://doi.org/10.1007/BF01211167} {\bibfield  {journal} {\bibinfo
  {journal} {Commun. Math. Phys.}\ }\textbf {\bibinfo {volume} {103}},\
  \bibinfo {pages} {599} (\bibinfo {year} {1986})}\BibitemShut {NoStop}%
\bibitem [{\citenamefont {Luscher}(2010)}]{Luscher:2010we}%
  \BibitemOpen
  \bibfield  {author} {\bibinfo {author} {\bibfnamefont {M.}~\bibnamefont
  {Luscher}},\ }\bibfield  {title} {\bibinfo {title} {Topology, the {W}ilson
  flow and the {HMC} algorithm},\ }\href {https://doi.org/10.22323/1.105.0015}
  {\bibfield  {journal} {\bibinfo  {journal} {PoS}\ }\textbf {\bibinfo {volume}
  {LATTICE2010}},\ \bibinfo {pages} {015} (\bibinfo {year} {2010})},\ \Eprint
  {https://arxiv.org/abs/1009.5877} {arXiv:1009.5877 [hep-lat]} \BibitemShut
  {NoStop}%
\bibitem [{\citenamefont {Alexandrou}\ \emph {et~al.}(2020)\citenamefont
  {Alexandrou}, \citenamefont {Athenodorou}, \citenamefont {Cichy},
  \citenamefont {Dromard}, \citenamefont {Garcia-Ramos}, \citenamefont
  {Jansen}, \citenamefont {Wenger},\ and\ \citenamefont
  {Zimmermann}}]{Alexandrou:2017hqw}%
  \BibitemOpen
  \bibfield  {author} {\bibinfo {author} {\bibfnamefont {C.}~\bibnamefont
  {Alexandrou}}, \bibinfo {author} {\bibfnamefont {A.}~\bibnamefont
  {Athenodorou}}, \bibinfo {author} {\bibfnamefont {K.}~\bibnamefont {Cichy}},
  \bibinfo {author} {\bibfnamefont {A.}~\bibnamefont {Dromard}}, \bibinfo
  {author} {\bibfnamefont {E.}~\bibnamefont {Garcia-Ramos}}, \bibinfo {author}
  {\bibfnamefont {K.}~\bibnamefont {Jansen}}, \bibinfo {author} {\bibfnamefont
  {U.}~\bibnamefont {Wenger}},\ and\ \bibinfo {author} {\bibfnamefont
  {F.}~\bibnamefont {Zimmermann}},\ }\bibfield  {title} {\bibinfo {title}
  {Comparison of topological charge definitions in lattice {QCD}},\ }\href
  {https://doi.org/10.1140/epjc/s10052-020-7984-9} {\bibfield  {journal}
  {\bibinfo  {journal} {Eur. Phys. J. C}\ }\textbf {\bibinfo {volume} {80}},\
  \bibinfo {pages} {424} (\bibinfo {year} {2020})},\ \Eprint
  {https://arxiv.org/abs/1708.00696} {arXiv:1708.00696 [hep-lat]} \BibitemShut
  {NoStop}%
\bibitem [{\citenamefont {Bhattacharya}\ \emph {et~al.}(2021)\citenamefont
  {Bhattacharya}, \citenamefont {Cirigliano}, \citenamefont {Gupta},
  \citenamefont {Mereghetti},\ and\ \citenamefont
  {Yoon}}]{Bhattacharya:2021lol}%
  \BibitemOpen
  \bibfield  {author} {\bibinfo {author} {\bibfnamefont {T.}~\bibnamefont
  {Bhattacharya}}, \bibinfo {author} {\bibfnamefont {V.}~\bibnamefont
  {Cirigliano}}, \bibinfo {author} {\bibfnamefont {R.}~\bibnamefont {Gupta}},
  \bibinfo {author} {\bibfnamefont {E.}~\bibnamefont {Mereghetti}},\ and\
  \bibinfo {author} {\bibfnamefont {B.}~\bibnamefont {Yoon}},\ }\bibfield
  {title} {\bibinfo {title} {Contribution of the {QCD} {$\Theta$}-term to the
  nucleon electric dipole moment},\ }\href
  {https://doi.org/10.1103/PhysRevD.103.114507} {\bibfield  {journal} {\bibinfo
   {journal} {Phys. Rev. D}\ }\textbf {\bibinfo {volume} {103}},\ \bibinfo
  {pages} {114507} (\bibinfo {year} {2021})},\ \Eprint
  {https://arxiv.org/abs/2101.07230} {arXiv:2101.07230 [hep-lat]} \BibitemShut
  {NoStop}%
\bibitem [{\citenamefont {van Baal}(1982)}]{vanBaal:1982ag}%
  \BibitemOpen
  \bibfield  {author} {\bibinfo {author} {\bibfnamefont {P.}~\bibnamefont {van
  Baal}},\ }\bibfield  {title} {\bibinfo {title} {Some results for {SU(N)}
  gauge fields on the hypertorus},\ }\href {https://doi.org/10.1007/BF01403503}
  {\bibfield  {journal} {\bibinfo  {journal} {Commun. Math. Phys.}\ }\textbf
  {\bibinfo {volume} {85}},\ \bibinfo {pages} {529} (\bibinfo {year}
  {1982})}\BibitemShut {NoStop}%
\bibitem [{\citenamefont {Brower}\ \emph {et~al.}(2003)\citenamefont {Brower},
  \citenamefont {Chandrasekharan}, \citenamefont {Negele},\ and\ \citenamefont
  {Wiese}}]{Brower:2003yx}%
  \BibitemOpen
  \bibfield  {author} {\bibinfo {author} {\bibfnamefont {R.}~\bibnamefont
  {Brower}}, \bibinfo {author} {\bibfnamefont {S.}~\bibnamefont
  {Chandrasekharan}}, \bibinfo {author} {\bibfnamefont {J.~W.}\ \bibnamefont
  {Negele}},\ and\ \bibinfo {author} {\bibfnamefont {U.~J.}\ \bibnamefont
  {Wiese}},\ }\bibfield  {title} {\bibinfo {title} {{QCD} at fixed topology},\
  }\href {https://doi.org/10.1016/S0370-2693(03)00369-1} {\bibfield  {journal}
  {\bibinfo  {journal} {Phys. Lett. B}\ }\textbf {\bibinfo {volume} {560}},\
  \bibinfo {pages} {64} (\bibinfo {year} {2003})},\ \Eprint
  {https://arxiv.org/abs/hep-lat/0302005} {arXiv:hep-lat/0302005} \BibitemShut
  {NoStop}%
\bibitem [{\citenamefont {Guo}\ \emph {et~al.}(2015)\citenamefont {Guo},
  \citenamefont {Horsley}, \citenamefont {Meissner}, \citenamefont {Nakamura},
  \citenamefont {Perlt}, \citenamefont {Rakow}, \citenamefont {Schierholz},
  \citenamefont {Schiller},\ and\ \citenamefont {Zanotti}}]{Guo:2015tla}%
  \BibitemOpen
  \bibfield  {author} {\bibinfo {author} {\bibfnamefont {F.~K.}\ \bibnamefont
  {Guo}}, \bibinfo {author} {\bibfnamefont {R.}~\bibnamefont {Horsley}},
  \bibinfo {author} {\bibfnamefont {U.~G.}\ \bibnamefont {Meissner}}, \bibinfo
  {author} {\bibfnamefont {Y.}~\bibnamefont {Nakamura}}, \bibinfo {author}
  {\bibfnamefont {H.}~\bibnamefont {Perlt}}, \bibinfo {author} {\bibfnamefont
  {P.~E.~L.}\ \bibnamefont {Rakow}}, \bibinfo {author} {\bibfnamefont
  {G.}~\bibnamefont {Schierholz}}, \bibinfo {author} {\bibfnamefont
  {A.}~\bibnamefont {Schiller}},\ and\ \bibinfo {author} {\bibfnamefont
  {J.~M.}\ \bibnamefont {Zanotti}},\ }\bibfield  {title} {\bibinfo {title} {The
  electric dipole moment of the neutron from 2+1 flavor lattice {QCD}},\ }\href
  {https://doi.org/10.1103/PhysRevLett.115.062001} {\bibfield  {journal}
  {\bibinfo  {journal} {Phys. Rev. Lett.}\ }\textbf {\bibinfo {volume} {115}},\
  \bibinfo {pages} {062001} (\bibinfo {year} {2015})},\ \Eprint
  {https://arxiv.org/abs/1502.02295} {arXiv:1502.02295 [hep-lat]} \BibitemShut
  {NoStop}%
\end{thebibliography}%
\end{document}